\documentclass[12pt]{iopart}

\usepackage{graphics}
\usepackage{graphicx}
\usepackage{epsfig}
\begin{document}

\title[MCMC Exploration of SMBHB Inspirals]{MCMC Exploration of Supermassive Black Hole
Binary Inspirals}

\author{Neil J. Cornish and Edward K. Porter}

\address{Department of Physics, Montana State University, Bozeman,
MT 59717, USA}

\begin{abstract}

The Laser Interferometer Space Antenna will be able to detect 
the inspiral and merger of Super Massive Black Hole Binaries (SMBHBs)
anywhere in the Universe. Standard matched filtering techniques can be used
to detect and characterize these systems. Markov Chain Monte Carlo (MCMC)
methods are ideally suited to this and other LISA data analysis problems
as they are able to efficiently handle models with large dimensions.
Here we compare the posterior parameter distributions derived by an MCMC
algorithm with the distributions predicted by the Fisher information matrix.
We find excellent agreement for the extrinsic parameters,
while the Fisher matrix slightly overestimates errors in the intrinsic parameters.
\end{abstract}


\submitto{\CQG}

\maketitle

\section{Introduction}

Super Massive Black Hole Binaries (SMBHBs) are likely to be the most powerful
sources of gravitational waves in the Universe. The Laser Interferometer
Space Antenna (LISA)~\cite{LISA} will be able to detect these systems out to the edge
of the visible Universe. By studying SMBHB inspirals we may gain insight into
the galaxy merger history and the role that black holes play in structure
formation. SMBHB inspirals and their subsequent mergers also provide fertile
ground for performing tests of general relativity~\cite{berti}.

To date, most data analysis development work for LISA has focused on Extreme
Mass Ratio Inspirals~\cite{gair} and the galactic confusion
problem~\cite{gclean,mcmc,genetic,tomographic}. SMBHB inspirals have received
relatively little attention, perhaps in part because they are not expected to
pose much of a challenge. The Markov Chain Monte Carlo (MCMC)
method has emerged as a leading algorithm for LISA data analysis.  The MCMC
method can efficiently explore large parameter spaces, perform model comparisons,
estimate instrument noise, while simultaneously providing error estimates for the
recovered parameters.  MCMC techniques have been applied
to ground based gravitational wave data analysis~\cite{christ}; a toy LISA problem~\cite{umstat};
and the extraction of multiple overlapping galactic binaries from simulated LISA data~\cite{mcmc}.
Here we make an initial foray into the SMBHB inspiral problem, and compare
the parameter recovery accuracy of an MCMC search to the predictions of the
Fisher information matrix.  Considerable work has been done on using the
Fisher information matrix to make predictions about LISA's resolving
abilities for SMBHBs~\cite{cutler98,hellings,hughes,Seto,alberto}, so it is
interesting to see how reliable those estimates might be.

We find that the Fisher matrix approach yields very good estimates for the
angular resolution and distance uncertainties, and that it
tends to overestimate the errors in the component masses and the
time of coalescence.

\section{The Model}
In this study we employ restricted post-Newtonian (PN) waveforms, which
neglect the higher order harmonics, and treat the
phase to 2-PN order~\cite{cutler98,biww}. The detector response is modeled using the
low frequency approximation~\cite{cutler98,rc}. The gravitational waveform for a
Supermassive black hole system consisting of two Schwarzschild black holes is described by
a 9-D parameter set, $\vec{x}=\{\ln(M_{c}),\ln(\mu),\theta, \phi, \ln(t_{c}),
\iota, \varphi_{c}, \ln(D_{L}), \psi\}$, where $M_{c}$ is the chirp mass, $\mu$ is
the reduced-mass, $(\theta,\phi)$ give the sky location, $t_{c}$ is the time-to-coalescence,
$\iota$ is the inclination of the orbital plane of the binary, $\varphi_{c}$ is the phase of
the gravitational wave at coalescence, $D_{L}$ is the luminosity distance and $\psi$ is the polarization
angle of the gravitational wave.  We will describe the parameters $D_{L}, \iota, \varphi_{c}, \psi$ as being
extrinsic, while all the rest will be described as being intrinsic~\cite{ben}.
We should mention that $(\theta,\phi)$ and $t_c$ would normally be classed as extrinsic, but they
become quasi-intrinsic due to the motion of the LISA observatory.

We focus our attention on a particular SMBHB system consisting of a $10^{7}-10^{6}\,M_{\odot}$ binary
system at $z=1$.
This gives corresponding values of $(M_{c},\mu, D_{L}) = (4.93\times10^{6}\,M_{\odot},
1.82\times10^{6}\,M_{\odot}, 6.63\,Gpc)$.  The other parameters are defined by $(\theta, \phi,
\iota, \varphi_{c}, \psi)=(1.325, 2.04, 1.02, 0.954, 0.658)$ radians.  We choose $t_{c} = 0.49$ years,
and set the time of observation to be $T_{obs}=0.5$ years. We used a sample cadence of 800 seconds.
Due to the the fact that the equations
describing the phase evolution of the wave break down before we reach the last stable circular
orbit (LSO), we terminate the waveforms at $R=7M$.  The source has a signal-to-noise ratio of $\sim$450.  

\begin{figure}[t]
\vspace*{0.4in}
\begin{center}
\epsfig{file=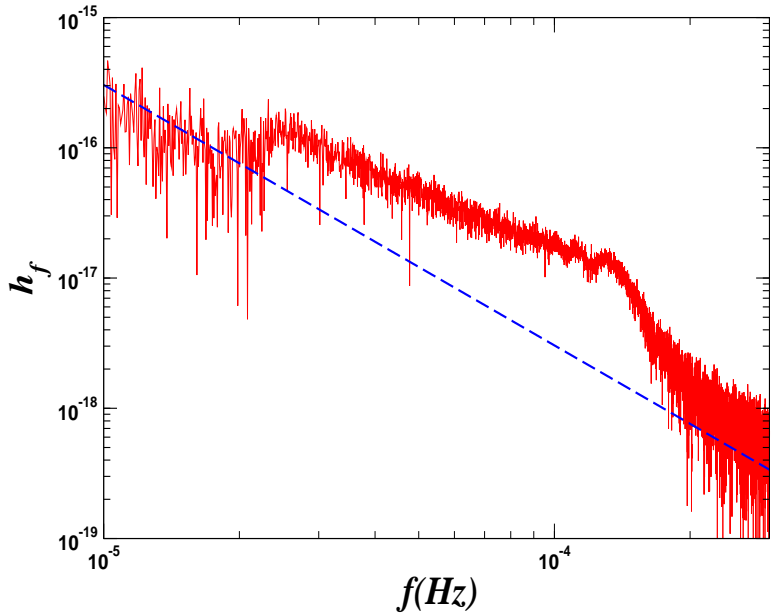, width=4in, height=3.2in}
\end{center}
\caption{A plot of the strain spectral density of the LISA response to a $10^{7}-10^{6}\,M_{\odot}$
binary at $z=1$. The dashed line indicates the RMS instrument noise level.}
\label{fig:spn}
\end{figure}

The one-sided noise spectral density for a LISA Michelson channel is given in the low frequency limit by
\begin{equation}
S_{n}(f)=\frac{1}{4L^{2}}\left[4 S_{n}^{pos}+16\frac{S_{n}^{accel}}{(2\pi f)^{4}}\right] , 
\end{equation}
where $L=5\times10^{6}$ kms is the arm-length for LISA,  $S_{n}^{pos} = 4\times10^{-22}\,m^{2}/Hz$
and $S_{n}^{accel} = 9\times10^{-30}\,m^{2}/s^{4}/Hz$ are the position and acceleration noise
respectively. Using this formula, we generate instrumental noise from a Gaussian distribution.
In Figure~\ref{fig:spn} we display the power spectrum of the LISA response to the system we are
investigating with instrumental noise from one channel of LISA.  Our analysis uses LISA as a
two channel detector, where the detectors are rotated by a factor of $\pi/2$ with respect to
one another~\cite{cutler98}. We also assume that LISA has a lower frequency cutoff of $10^{-5}$ Hz.
\section{The MCMC approach}
In the Bayesian framework of data analysis, the goal is to calculate the posterior probability
density functions for the parameters of the model, given the parameter priors and the data. 
However, this is extremely difficult, if not impossible, to do analytically.  The advantage of
the MCMC method is that it allows one to map out the posterior numerically.  Our MCMC approach
is implemented by using a Metropolis-Hastings algorithm to map out the posterior distributions
of the 9-D parameter space for Super Massive black holes.  The method works as follows: We assume
that we have already run a search chain that has found the system we are looking for. We use the
true parameter values as the starting point for our chain $\vec{x}$. The algorithm then suggests
a jump to a new point $\vec{y}$ using a proposal distribution $q(^{.}|\vec{x})$. We evaluate the
Hastings ratio 
\begin{equation}
H = \frac{\pi(\vec{y})p(s|\vec{y})q(\vec{x}|\vec{y})}{\pi(\vec{x})p(s|\vec{x})q(\vec{y}|\vec{x})},
\end{equation}
and accept the jump with probability $\alpha = {\rm min}(1,H)$, otherwise the chain stays
at its current position. Here $\pi(\vec{x})$ are the priors of the parameters and $p(s|\vec{x})$ is
the likelihood. We use uniform priors for all the parameters. The parameters
$\ln M_c$, $\ln \mu$, $\ln t_c$ and $\ln D_L$ were taken to be uniform in the range
$(-\infty,\infty)$, $\cos \theta$ and $\cos \iota$ were taken to be uniform in the
range $[-1,1]$, $\phi$ and $\varphi_c$ were taken to be uniform in the range $[0,2 \pi]$,
and $\psi$ was taken to be uniform in the range $[0,\pi]$. The log-likelihood for a template $h(\vec{x})$
given a signal $s$ was assumed to have the form
\begin{equation}
\ln p\left(s |\vec{x} \right) = -\frac{1}{2} \langle s -h(\vec{x})  \vert  s -h(\vec{x}) \rangle.
\end{equation}
Here the angular brackets define the standard noise weighted inner product.  In order to 
achieve a healthy acceptance rate for the proposed jumps, we use a proposal distribution
given by a multivariate normal distribution that is the product of independent normal distributions
in each of the 9 eigendirection of the Fisher matrix, $\Gamma_{ij}(\vec{x})$.  The Fisher information matrix
describes the expectation value of the curvature of the log likelihood function at maximum likelihood:
\begin{equation}
\Gamma_{ij}(\vec{x}_{\rm ML}) = \overline{\partial_i \partial_j \ln p\left(s |\vec{x}_{\rm ML} \right)} = 
\langle \partial_i h(\vec{x}_{\rm ML})  \vert  \partial_j h(\vec{x}_{\rm ML}) \rangle \, .
\end{equation}
Here the over line indicates the expectation value. We used a generalized notion of the Fisher
matrix by employing the definition $\Gamma_{ij}(\vec{x}) 
= \langle \partial_i h(\vec{x})  \vert  \partial_j h(\vec{x}) \rangle$ for points $\vec{x}$ away from
maximum likelihood. The standard deviation in
each eigendirection of $\Gamma_{ij}(\vec{x})$ was set to equal $\sigma_{i} = 1/\sqrt{DE_{i}(\vec{x})}$, where
$D=9$ and $E_{i}(\vec{x})$ is the corresponding eigenvalue.  The factor of $1/\sqrt{D}$ ensures that the typical
jumps are $\sim 1 \sigma$ for the full multivariate normal distribution. In principle this choice
of proposal distribution should yield a $\sim 69 \%$ acceptance rate for the proposed jumps. In practice
we found an acceptance rate of $\sim 33\%$ for the system being studied. The lower acceptance rate is due to
the Fisher matrix slightly overestimating the uncertainties in one of the eigendirections,
and to a slight error in the estimate of the orientation of the corresponding eigendirection.

\section{Results}

\begin{figure}[t]
\vspace*{0.6in}
\begin{center}
\epsfig{file=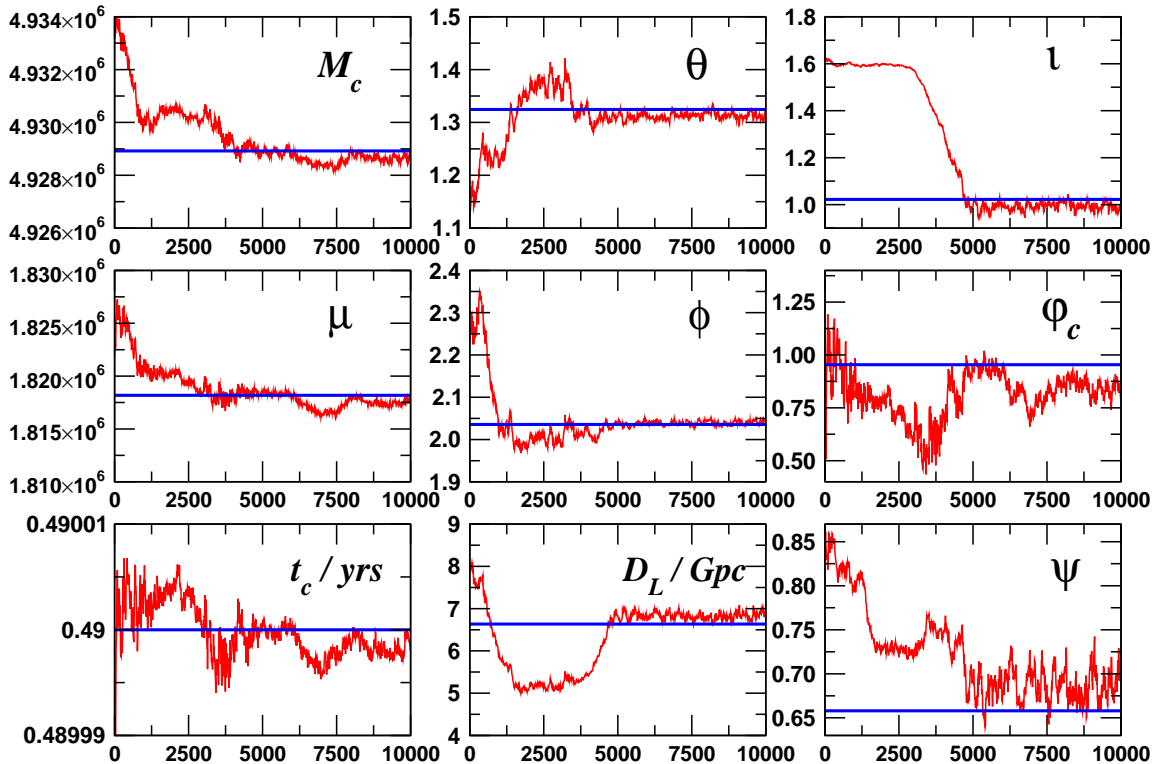, width=6in, height=4in}
\end{center}
\vspace*{0.1in}
\caption{A plot of the MCMC chains for the 9 parameters describing the SMBHB binary during the
initial burn-in phase.  The straight line in each panel denotes the true parameter value.}
\label{fig:burn}
\end{figure}

We started the chain with parameter values close to their true values, but offset by
$\Delta m_1 /m_1 = -0.2 \%$, $\Delta m_2 /m_2 = +0.3 \%$, $\Delta z = +0.2$,
$\Delta t_c/t_c = -2 \times 10^{-5}$, $\Delta \theta = -0.2$, $\Delta \phi = +0.2$,
$\Delta \iota = +0.6$, $\Delta \psi = +0.2$, and $\Delta \phi_c = -0.4$. The offset in
each parameter was chosen to be $\sim$tens of standard deviations away from the true
source parameters (as estimated by the Fisher Information Matrix). In Figure~\ref{fig:burn}
we see that the chain underwent a burn-in phase during the first 10,000 steps, before settling
in close to the true source parameters. In repeated trials we found that the chain always settled
in to the same region of parameter space after the burn-in was completed. When the chain was
started off far from the true parameters the burn-in time became prohibitively long, so we
do not recommend that the current algorithm be used for blind searches. We have developed
a non-Markovian variant of the algorithm that can efficiently handle the search phase~\cite{search}.
The posterior distribution functions derived after the full search are identical to those
found in the current paper.

\begin{figure}[t]
\vspace*{0.6in}
\begin{center}
\epsfig{file=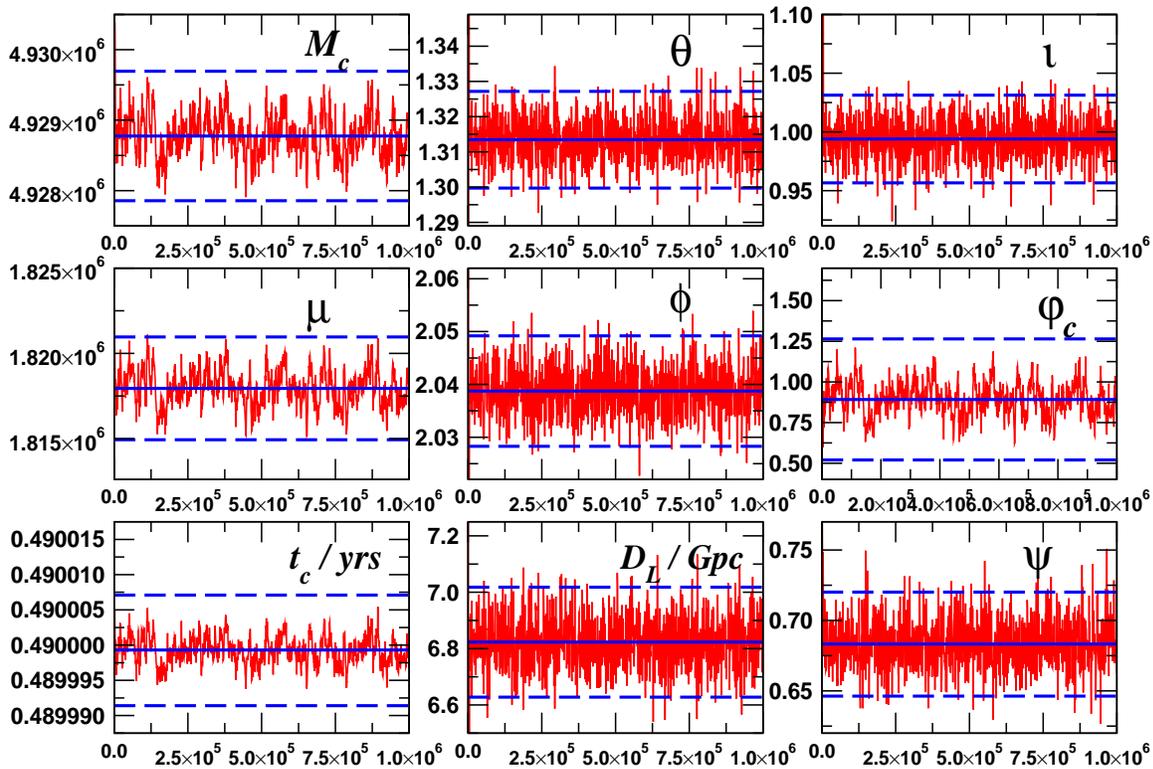, width=6in, height=4in}
\end{center}
\vspace*{0.1in}
\caption{A plot of the MCMC chains for the 9 parameters describing the SMBHB binary.  The straight
line in each panel denotes the mean of the chain as calculated from the $8 \times 10^{6}$ points.  The
dashed lines are the $2-\sigma$ predictions from the Fisher matrix evaluated at the mean of
each chain.}
\label{fig:chains}
\end{figure}

Following the burn-in, the chain was run for another $8 \times 10^6$ points to explore the posterior
distribution function. This took 8 days on a single 2 GHz processor. The first million points of the
exploration phase of the chain are shown in Figure~\ref{fig:chains}.
We see that the MCMC chain moves around well, and show clear evidence of the high degree of
correlation between the parameters $M_c, \mu, t_c$ and $\varphi_c$. The straight, solid line in
each panel of Figure~\ref{fig:chains} denotes the mean value of the parameter, as
calculated from the chain. Taking this mean value, we then calculated the standard deviation as predicted by the
Fisher matrix at that point.  This $1-\sigma$ error is given by
\begin{equation}
\sigma_{i} = \sqrt{C_{ii}}, 
\end{equation}
where $C_{ij}=(\Gamma_{ij})^{-1}$.  The two dashed lines in each panel denote the $\pm 2\sigma$
errors in each of the parameters.

\begin{figure}[t]
\vspace*{0.6in}
\begin{center}
\epsfig{file=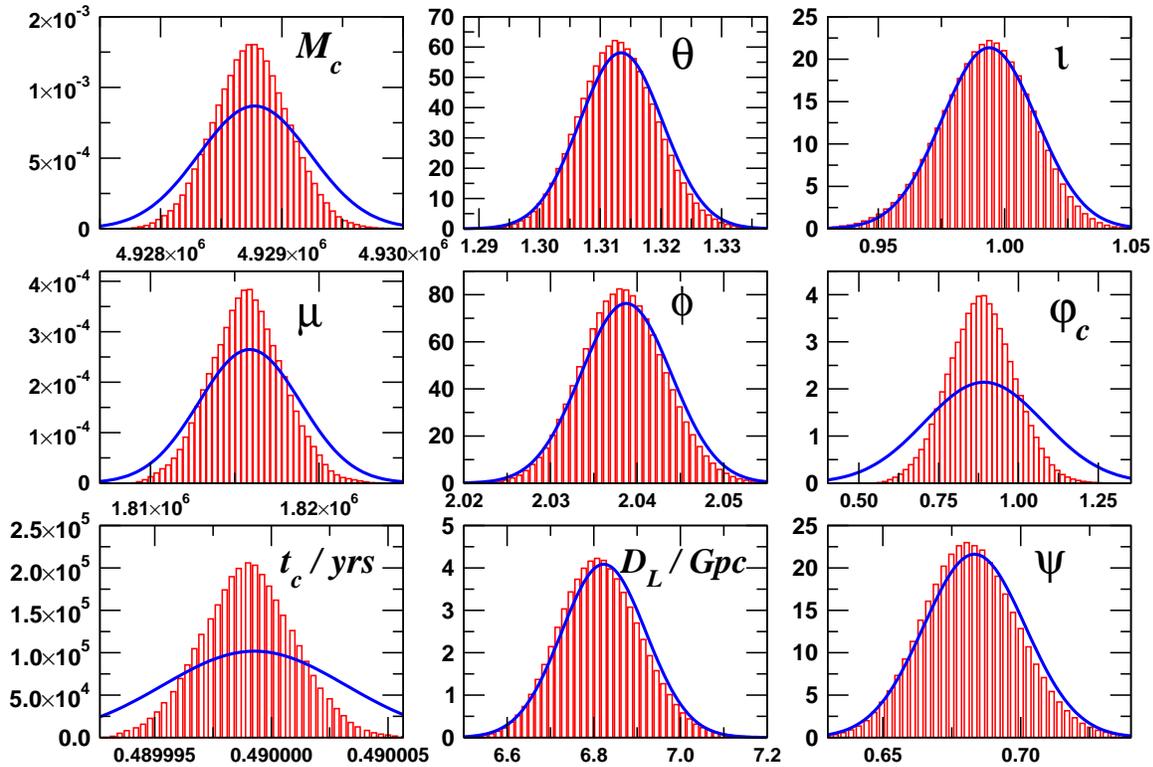, width=6in, height=4in}
\end{center}
\vspace*{0.3in}
\caption{A plot of the marginalized 1-d histograms for each of the nine parameters.  We see
that most of the angular variables agree almost exactly with the prediction of the Fisher matrix.
However, the remaining four parameters, which are all highly correlated, differ from the
Fisher matrix prediction for the posterior.}
\label{fig:histograms}
\end{figure}

\begin{figure}[t]
  \vspace{9pt}

  \centerline{\hbox{ \hspace{0.0in} 
    \epsfxsize=3.0in
    \epsffile{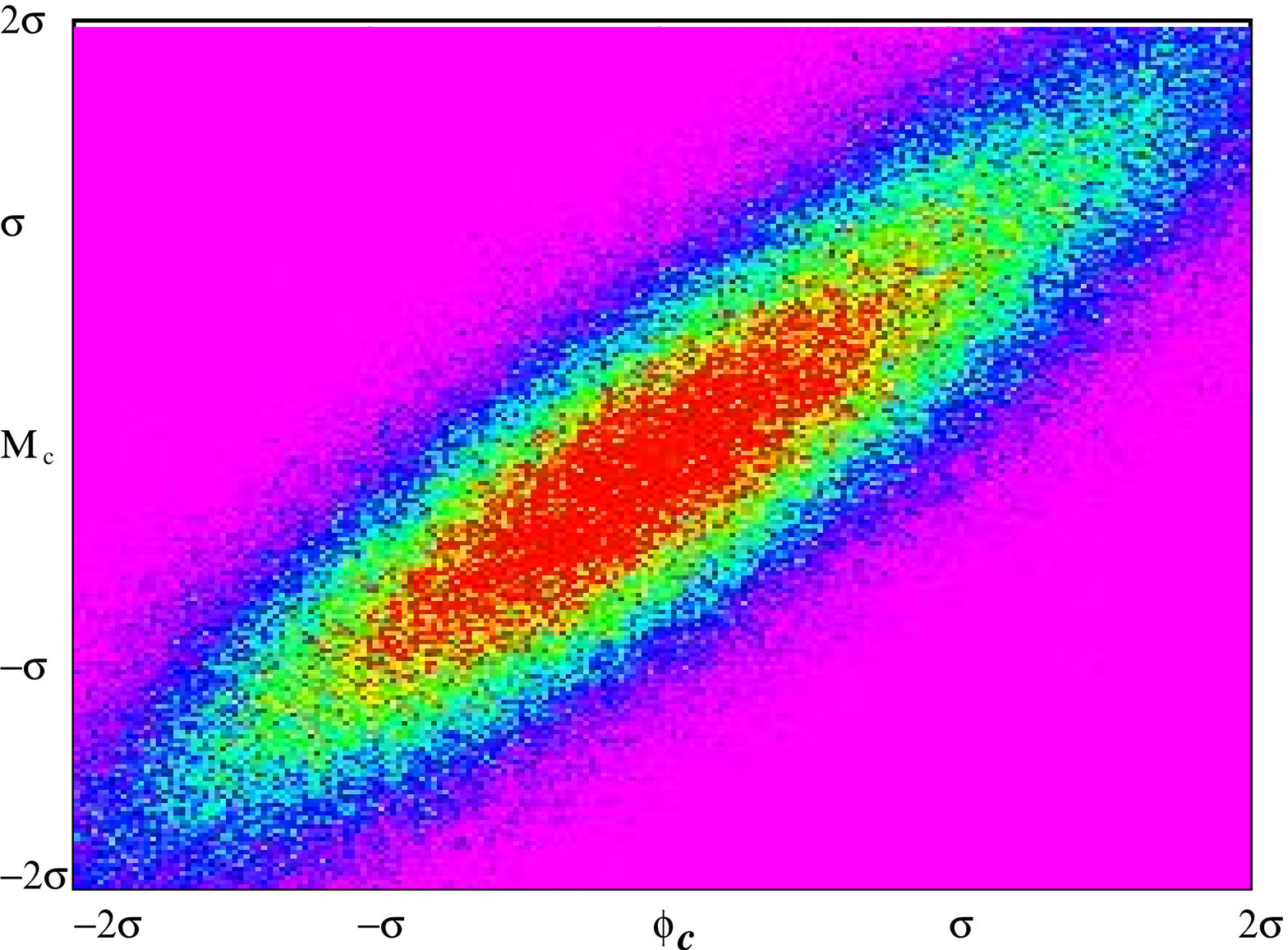}
    \hspace{0.15in}
    \epsfxsize=3.0in
    \epsffile{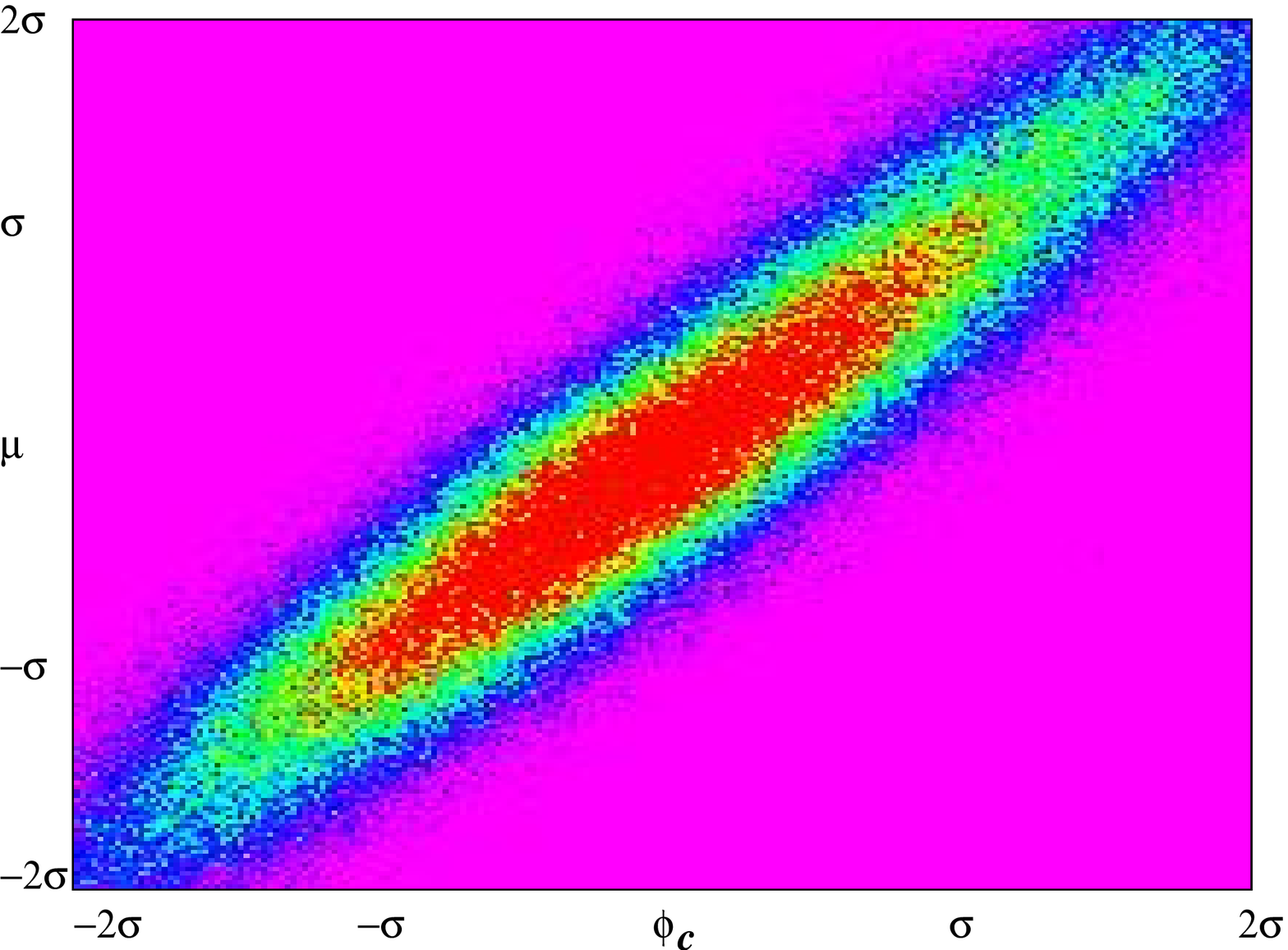}
    }
  }

  \vspace{9pt}

  \centerline{\hbox{ \hspace{0.0in}
    \epsfxsize=3.0in
    \epsffile{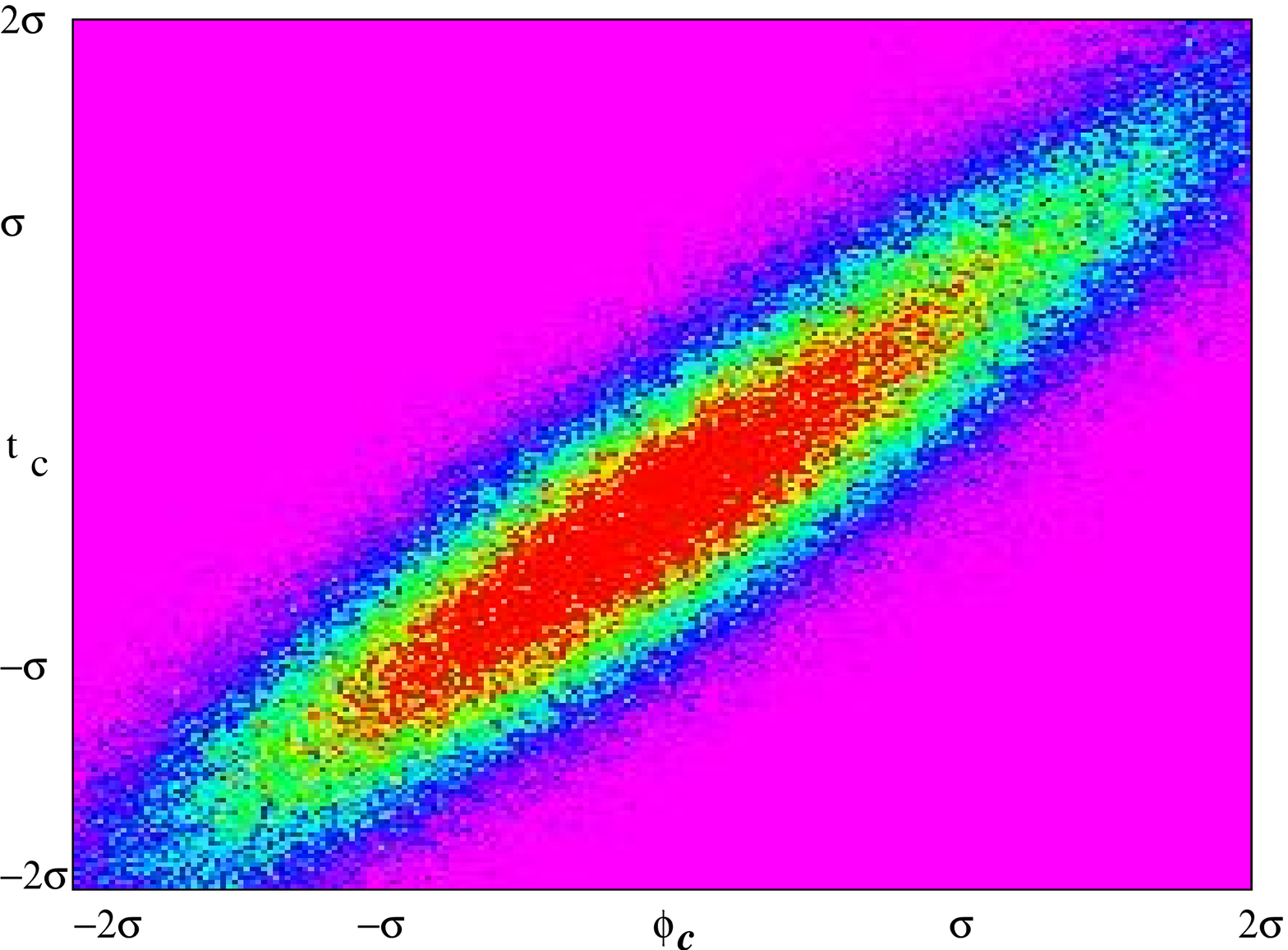}
    \hspace{0.15in}
    \epsfxsize=3.0in
    \epsffile{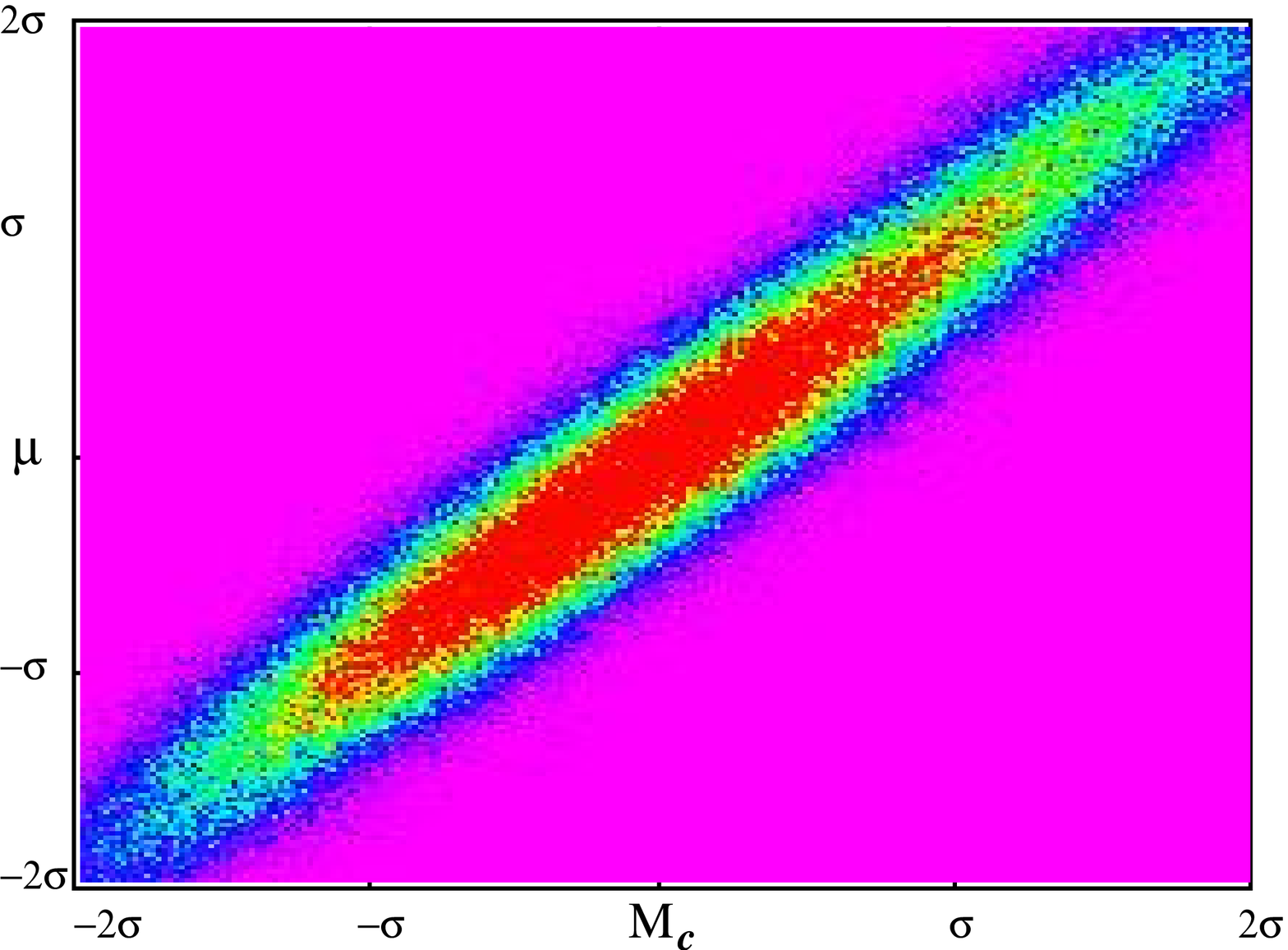}
    }
  }

  \vspace{9pt}

  \caption{This figure showing the 2-D marginalized histograms demonstrates the large correlations
between the parameters $(M_{c},\mu,t_{c}, \varphi_{c})$.}
  \label{fig:2dhistograms}

\end{figure}

In Figure~\ref{fig:histograms} we have plotted the 1-D marginalized histograms for each of the
parameters.  The solid line is the Fisher matrix prediction for the error at the mean of each
chain.  We can see that most of the extrinsic parameter histograms match the Fisher prediction
almost perfectly.  This is also the case for the $(\theta,\phi)$ parameters.  However, we can
see that the histograms for $(M_{c},\mu,t_{c}, \varphi_{c})$ all differ from the prediction
of the Fisher matrix. We attribute the error in the Fisher matrix prediction to the high
degree of correlation between these parameters. In order to see just how correlated these parameters are,
in Figure~\ref{fig:2dhistograms} we plot the 2-D marginalized histograms for the
combinations $(M_{c},\varphi_{c})$, $(\mu,\varphi_{c})$, $(t_{c},\varphi_{c})$,
and $(M_{c},\mu)$. We found that the Fisher matrix has a small eigenvalue in
the $(M_{c},\mu,t_{c}, \varphi_{c})$ direction which dominates the contribution
to the proposed jumps in these parameters. The Fisher matrix tends to under estimate
the eigenvalue in this direction, and it is also slightly off in predicting the corresponding
eigendirection. Similar behavior was seen in all the examples we have looked at, which suggests
that the Fisher matrix estimates for the uncertainties in the component masses found in the
literature may be systematically high by about a factor of two.

In a closely related study that appears in these same proceedings, Wickham, Stroeer \& Vecchio~\cite{WSV}
use a Reversible Jump MCMC routine to study the posterior parameter distributions of a SMBHB system. They
used a simplified model of the SMBHB waveform, which removes the reduced mass from the parameter set.
In contrast to our findings, they found that the Fisher matrix significantly underestimated the uncertainties
in many of the model parameters. We do not know why our results are so different, but we do not think it is
due to the differences in our MCMC implementations. The Reverse Jump method should yield the same results
as our standard MCMC implementation when applied to models with fixed numbers of parameters.

\section{Discussion}

We have found that it is possible to construct a simple MCMC sampler for studying the
posterior distribution functions for SMBHB inspirals in the LISA data streams. The marginalized posterior
distributions recovered from the Markov chains are in good to fair agreement with the Fisher matrix
predictions. In another work~\cite{search}, we have developed an advanced MCMC search routine that
is capable of performing a blind search of the LISA data for SMBHB inspirals.

\end{document}